\documentclass[twocolumn,preprintnumbers,amsmath,amssymb,aps,prl,reprint]{revtex4}
\usepackage{graphicx}

\begin{document}

\title{
Reversible to Irreversible Transitions in Periodically Driven Skyrmion Systems 
} 
\author{
B. L. Brown$^{1,2}$, C. Reichhardt$^{1}$, and C. J. O. Reichhardt$^{1}$
} 
\affiliation{
$^1$ Theoretical Division and Center for Nonlinear Studies,
Los Alamos National Laboratory, Los Alamos, New Mexico 87545, USA\\ 
$^2$ Department of Physics, Virginia Tech, Blacksburg, Virginia 24061-0435, USA\\ 
} 

\date{\today}
\begin{abstract}
We examine skyrmions driven periodically over random quenched disorder 
and show that there is a transition from reversible motion to a state in which
the skyrmion trajectories are chaotic or irreversible. 
We find that the characteristic time required
for the system to organize into a steady reversible or irreversible state
exhibits a power law divergence
near a critical ac drive period, with the same exponent as that observed for
reversible to irreversible transitions in periodically sheared colloidal
systems, suggesting that the transition can be described as
an absorbing phase transition in the directed percolation
universality class.
We compare our results to the
behavior of an overdamped system
and show that the Magnus term enhances the
irreversible behavior
by increasing the number of dynamically accessible orbits.
We discuss the implications
of this work for skyrmion applications involving
the long time repeatable dynamics of dense skyrmion arrays.
\end{abstract}
\maketitle

A great variety of many body systems exhibit 
nonequilibrium phases under an applied drive \cite{1},
including assemblies of colloidal particles \cite{2,3},
vortices in type-II superconductors \cite{4,5}, sliding friction \cite{6}, 
active matter \cite{7}, dislocation motion \cite{8}, and geological systems \cite{9}.
Despite the ubiquity of these systems,
it is often experimentally difficult to find a protocol
that will produce distinct nonequilibrium phases separated
by clear transitions,
and  in many cases it is not even clear
what order parameter or measures should be used.

In 2005 Pine {\it et al.} \cite{10} examined a fairly
simple system of a dilute assembly of colloidal particles 
in an overdamped medium subjected to periodic shearing.
They introduced a new stroboscopic measure
in which the positions of the particles at the end of each
shear cycle are compared
to the positions at the end of the previous cycle.
Pine {\it et al.} found that as a function of the distance over which the particles
are sheared,
there is a transition from a reversible state, in which all particles return
to the same positions after each cycle,
to an irreversible state,
in which the particles do not return to the same positions
and exhibit chaotic dynamics with a long time diffusive behavior.
The threshold shearing distance decreases
as the density of particles increases due to an increase in the frequency
of particle-particle collisions.
In further work on this system,
Corte {\it et al.} \cite{11} found that
the initial motion is always irreversible but that over time the
particles organize into a steady state that is either reversible or
irreversible depending on the shearing distance.
The number of shearing cycles required to reach a steady state
diverges as a power law at a critical point, suggesting that
the reversible-irreversible transition is an example of a nonequilibrium
phase transition.  The power law exponents are consistent with
those expected for an absorbing phase transition in the
directed percolation
universality class \cite{11,12,13}.
In the reversible regime,
all the fluctuations are lost
and the system is absorbed into a state which it cannot escape. 
More recent work has shown
that the periodically sheared colloidal system
organizes into a reversible state
in which particle-particle collisions no longer occur
and large-scale density fluctuations in the particle positions are suppressed,
giving hyperuniform order rather than a truly random state
\cite{14,15,16,17}. 
Similar reversible to irreversible transitions
have been observed in a wide range
of other periodically driven systems that are much more strongly interacting
than the colloidal
particles, such as granular matter \cite{18,19,20,21}, dislocations \cite{22,23},
amorphous solids \cite{24,25,26,27}, polycrystalline solids \cite{28},
charged colloids \cite{29}, and vortices in type-II superconductors \cite{30,31,32}.
The dynamics of most of these systems
is overdamped, and little is known about how nondissipative
dynamics would affect
a reversible to irreversible transition.  

An important example of nondissipative interacting particles is
skyrmions in chiral magnets.  Discovered in 2009, magnetic skyrmions are
nanometer-sized spin textures
that have many similarities to
vortices in type-II superconductors,
in that they form
a triangular lattice
in the absence of quenched disorder \cite{33,49,50}
and undergo driven motion when subjected to
an applied current \cite{33,51,52,53,54,55,56,57,58}. 
A key difference between skyrmions and the systems described above
is that the skyrmion dynamics includes  
a pronounced nondissipative Magnus term \cite{33,51,53}.
The Magnus term generates velocity components that are perpendicular to the
net force experienced by the skyrmion,
unlike 
the damping term which aligns the skyrmion velocity with the external forces.
In
the absence of quenched disorder,
under an applied drive
the skyrmions move at an angle
to the driving direction
called the intrinsic skyrmion Hall angle $\theta_{sk}^{\rm int}=\tan^{-1}(\alpha_m/\alpha_d)$,
given by the
ratio of the Magnus term $\alpha_m$ to the dissipative term $\alpha_d$ \cite{33}.
When the Magnus term is zero, $\theta_{sk}^{\rm int} = 0^\circ$.
The observed skyrmion Hall angle $\theta_{sk}$ has been shown to
become drive dependent in the presence of pinning due to a side jump
of the skyrmions produced by the Magnus term as the skyrmions move through
the pinning sites, which decreases in magnitude
as the skyrmion velocity increases,
giving a saturation to $\theta_{sk}=\theta_{sk}^{\rm int}$ at high drives
\cite{67,69,73,75}.
Continuum-based simulations of skyrmions confirm that
the drive dependence of the skyrmion Hall angle
exists when disorder is present and is absent in
the disorder-free limit \cite{57,76}.
In skyrmion experiments, after the skyrmions depin they enter a
flowing phase in which $\theta_{sk}$ is directly observed
to increase with drive before saturating at high drives
\cite{77,N}. 
A similar increase in $\theta_{sk}$
as a function of ac drive has also been found experimentally \cite{78}.

Skyrmions can potentially be used for a variety of applications
similar to those proposed for magnetic
domain walls,
where the smaller size and higher mobility
of the skyrmions give them
numerous advantages over domain wall systems
\cite{79}.
Many of the applications require
the motion of the skyrmions to be reversible,
and experiments involving ac drives have shown
that the dynamics of isolated skyrmions can remain reversible
over a large number of drive cycles
\cite{78}.
For applications
in which it is necessary for
dense arrays of interacting skyrmions
to maintain reversible motion
over many cycles,
it is important to
develop an understanding of
the onset
of periodic reversible behaviors
and to characterize the irreversible behaviors
as a function of the net displacements of the skyrmions
under an ac drive.

In this work we employ a periodic driving protocol to
a particle-based model of a skyrmion assembly interacting with random
disorder.
We characterize the system by analyzing the
net displacement of the skyrmions
between the beginning and end of each drive cycle,
where a fully reversible state corresponds to a net displacement of zero.
By varying the driving period, which changes the distance $d$ that an
isolated skyrmion in the absence of disorder can travel during a single
drive cycle,
we find that there
is a well defined
critical value $d_c$ below which the system reaches a steady reversible state
and above which a steady irreversible state emerges.
Near $d_c$ we find
a characteristic time scale $\tau$
for the system to reach a steady state,
where $\tau \propto |d - d_{c}|^{-\nu}$  with exponent $\nu \approx 1.3$. 
The divergence of $\tau$ at the
reversible-irreversible transition is
similar to what is observed in
the periodic shearing of dilute colloids \cite{11}          
and jammed solids \cite{25,27} as a function of increasing $d$,
and the power law exponent is also similar, suggesting that
$d_{c}$ is a critical point
separating
an absorbing reversible state from a fluctuating
irreversible state
in the directed percolation universality class \cite{12}.
In the overdamped limit, which represents
strongly damped skyrmions as well as vortices in type-II 
superconductors,
we observe a similar power law time scale divergence
in the reversible regime,
but we find evidence that there are
two distinct transitions
in the irreversible regime.
In the first transition,
the motion of the particles becomes
irreversible in the direction parallel to the drive when a
dynamically reordered smectic state appears, and in the second
transition, the motion becomes irreversible both parallel and
perpendicular to the drive.

\section{Simulation}
We simulate $N=245$
particles
in a two-dimensional (2D) sample of size $\frac{2}{\sqrt{3}}L\times L$ 
with periodic boundary conditions in the $x$ and $y$ directions.
We use a modified Thiele equation
for the particle-particle interactions
as in previous works \cite{67,73,75,80,81},
with ${\bf F}^{ss}_{i}=\sum^{N}_{i\ne j}F^{ss}_{0}K_{1}(R_{ij})\hat{\bf R}_{ij}$, where
$F_{0}^{ss}=1$,
$K_{1}$ 
is a modified Bessel function, $R_{ij}=|{\bf R}_{i}-{\bf R}_{j}|$ is the distance between 
particles $i$ and $j$, and $\hat{\bf R}_{ij}=({\bf R}_{i}-{\bf R}_{j})/R_{ij}$.
The interaction force
becomes very weak for $R_{ij}>1$ and we take interactions between particles $i$ and $j$ with $R_{ij}>7$ to be negligible.
We model the quenched disorder as $N_{p}=113$ randomly placed
nonoverlapping harmonic pinning sites with radius $r_{p}=0.3$ 
such that ${\bf F}^{p}_{i}=\sum^{N_{p}}_{k}  F^{p}_{0}(R_{ik}/r_{p})\Theta (r_{p}-R_{ik})\hat{\bf R}_{ik}$ where $F^{p}_0=1.5$,
$\Theta$ is the Heaviside step function and
$R_{ik}$ is the distance between particle $i$
and pinning site $k$.
The particles are driven by a periodic current applied to the sample which we model as a square wave with period $T$: ${\bf F}^{AC}=F^{AC}_{0}$sgn$(\sin(2\pi t/T))\hat{\bf x}$,
where $F^{AC}_{0}=1.3$.

The Langevin equations of motion are as follows:
\begin{equation}
  \label{EQ_1}
  \alpha_{d}{\bf v}_{i} + \alpha_{m}\hat{\bf z}\times{\bf v}_{i} = {\bf F}_{i}^{ss} + {\bf F}_{i}^{p} + {\bf F}^{AC},
\end{equation}
where ${\bf v}_{i}$ is the velocity of particle $i$,
$\alpha_{d}$ is the damping coefficient,
and $\alpha_{m}$ determines the strength of the Magnus term.
We use a standard fourth-order Runge-Kutta method to integrate Eq.~(\ref{EQ_1})
with a time step of $\Delta t=0.05$. 
In skyrmion systems, the nondissipative Magnus term is important to the dynamics, but for other systems 
such as vortices in type-II superconductors, the Magnus force is negligible and the dynamics 
are described by Eq.~(\ref{EQ_1}) in the overdamped limit $\alpha_{m}\rightarrow 0$ \cite{BLA94}. 
In this work we focus on two cases:
the skyrmion limit with $\alpha_{m}/\alpha_{d}=1$
and the overdamped limit with $\alpha_{m}/\alpha_d=0$.
In order to
facilitate comparisons of the skyrmion and overdamped limits we fix $\alpha_{m}^{2} +\alpha_{d}^2=1$.
We report the magnitude of the ac drive in terms of
the displacement $d$ of an isolated particle in the absence of disorder
during half of a drive cycle,
$d=\frac{F^{AC}_{0}T}{2}$.
The particles are initialized in 
a triangular lattice
and are 
allowed to relax under the influence of the quenched disorder.
Once the relaxation is complete, we apply the periodic driving force
and measure the mean square displacement,
$R^{2}_{\alpha}(n)$, in the $x$ and $y$ directions as follows:
\begin{equation}
  \label{EQ_2}
  R^{2}_{\alpha}(n) = \Big\langle \frac{1}{N} \sum^{N}_{i}[(\tilde{\bf R}_{i}(nT)-{\bf R}_{i}(0)) \cdot \hat{\alpha}]^{2}  \Big\rangle,
\end{equation}
where $\alpha=(x,y)$, $n$ is the number of drive cycles over which the
displacement is measured, $\tilde{\bf R}_i$ is the absolute position of particle $i$
without reflection back into the periodic box,
and the brackets
indicate an average over different quenched disorder realizations.
After a transient time, the system
reaches a steady state
characterized by either reversible or irreversible flow
depending on the drive period.
The particle trajectories in the irreversible regime
are chaotic and $R^{2}_{\alpha}$ behaves as
an anisotropic random walk, while in
the reversible regime $R^2_{\alpha}$ approaches a constant value.
We also measure the fraction of active particles, $F_{a}(n)$, as a function of the number of cycles. A particle is
defined to be `active'
if it does not return to a circular region of
radius $r_{a}=2r_{p}$ centered on its position at the end of the last drive cycle, giving
\begin{equation}
  F_{a}(n)=\Big\langle\frac{1}{N}\sum_{i}^{N}\Theta(|\tilde{\bf R}_{i}(nT)-\tilde{\bf R}_{i}((n-1)T)|-r_{a})\Big\rangle.
  \label{EQ_3}
\end{equation}
In reversible flow, there are no active particles and $F_a \rightarrow 0$, while
$F_a$ approaches a finite value when the flow is irreversible.

\section{Results}

\begin{figure}
  \center
  \includegraphics[width=\columnwidth]{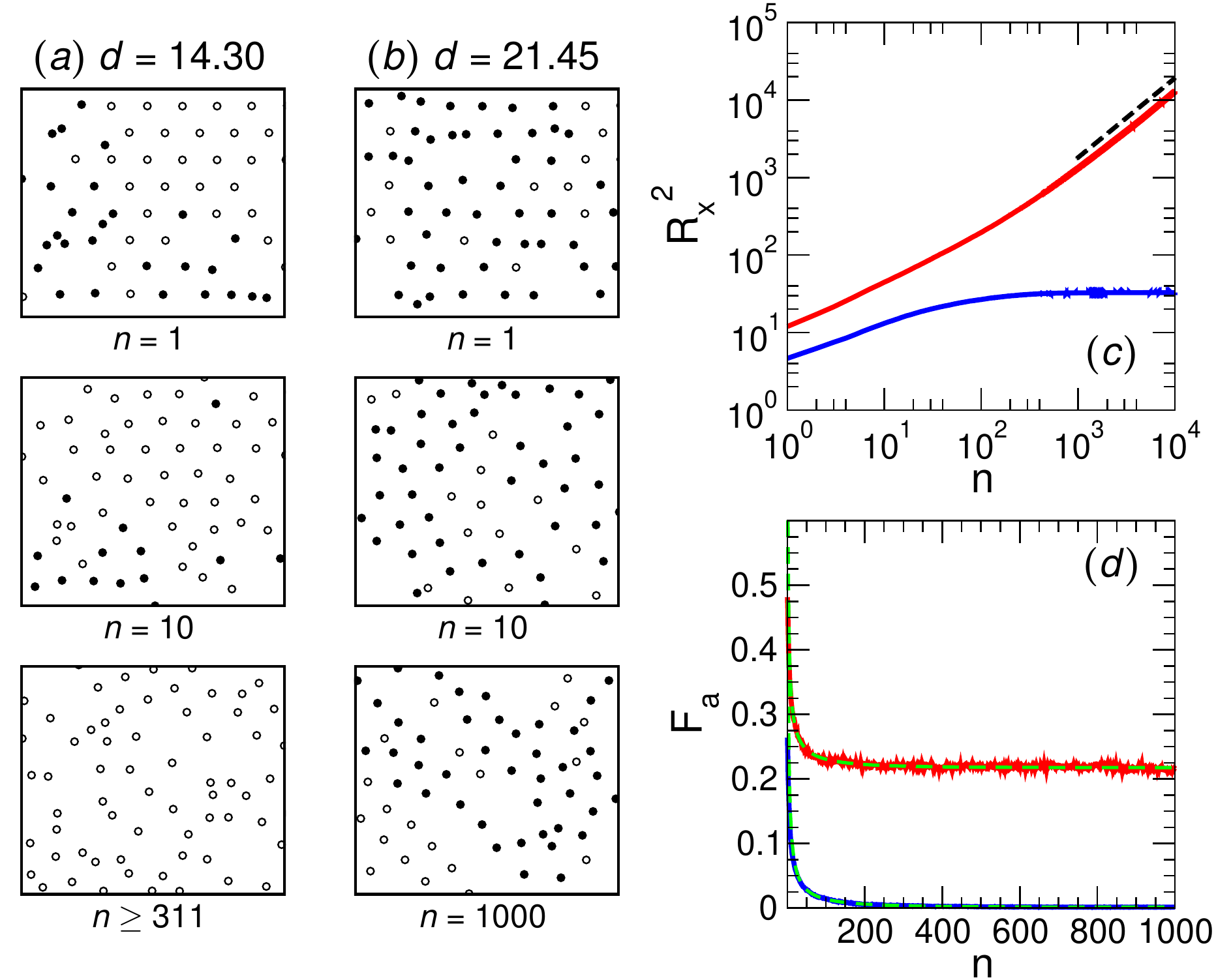}
  \caption{(a,b) Images of particle positions in a portion of the sample
    in the skyrmion limit with $\alpha_m/\alpha_d=1$.  Filled circles are `active'
    particles that do not return to the same position after each drive cycle, and
    open circles are inactive particles that move reversibly.
    (a) A reversible state at $d=14.30$ after $n=1$, 10, and 311 cycles, from top
    to bottom.
    (b) An irreversible state at $d=21.45$ after $n=1$, 10, and 1000 cycles, from
    top to bottom.
    (c) The mean square particle displacement in the $x$ direction, $R_x^2$, vs
    cycle number $n$ for $d=14.30$ (blue) and $d=21.45$ (red).
    The dashed line is a fit to $R_x^2 \propto n$.
    (d)
    The fraction $F_a$ of active particles vs $n$ for $d=14.30$ (blue) and $d=21.45$ (red).
    The dashed lines
    are fits to Eq.~(\ref{EQ_4}).}
  \label{FIG_1}
\end{figure}

Under small displacements, the particle trajectories become reversible after
a transient time interval of reorganization, and 
the system returns to the same configuration at the end of each
drive cycle.
Figure \ref{FIG_1}(a,b) shows snapshots of a portion of the system in the
skyrmion limit for reversible and irreversible flows.
For the reversible case with $d=14.30$ in Fig.~\ref{FIG_1}(a),
the system
reaches an absorbed completely
reversible state after $n=311$ cycles of the drive. 
As $d$ increases, we find
a transition to irreversible flow,
illustrated for $d=21.45$ in Fig.~\ref{FIG_1}(b),
in which the motion 
of the particles becomes chaotic and each active particle
moves further away from
its initial location
after each cycle.
Here the fraction $F_a$ of active particles remains finite since
the system is unable to find a reversible configuration.
In Fig.~\ref{FIG_1}(c)
we plot the mean square displacement parallel to the drive, $R^{2}_{x}$,
versus $n$ for the reversible and irreversible states.
At $d=14.30$ when the motion is reversible, $R^{2}_{x}$ 
approaches a constant value,
while for $d=21.45$ when the system is irreversible,
$R^{2}_{x}$ increases linearly with time at long times,
consistent with a one-dimensional
random walk.
This result is in agreement with Ref.~\cite{30},
where similar behavior is observed for superconducting vortices in the
irreversible flow regime.
In Fig.~\ref{FIG_1}(d), we plot $F_a$ versus $n$, which goes to zero when
$d=14.30$ in the reversible state, and saturates to a finite value
for $d=21.45$ in the irreversible state.
Following Cort\'{e} et al. \cite{11}, we find that the active fraction is well fit by the following relaxation function:
\begin{equation}
  F_{a}=(F_{a}^{0} - F_{a}^{\infty})e^{-t/\tau}t^{-\alpha} + F_{a}^{\infty} ,
  \label{EQ_4}
\end{equation}
where $F_{a}^{0}$ and $F_{a}^{\infty}$ are the initial and steady
state values of $F_{a}$, respectively,
and $\tau$ is the characteristic time at which the
relaxation crosses over from an exponential decay
to a power-law behavior.
The dashed lines in Fig.~\ref{FIG_1}(d)
indicate fits of $F_{a}$
to Eq.~(\ref{EQ_4}).

\begin{figure}
  \center
  \includegraphics[width=\columnwidth]{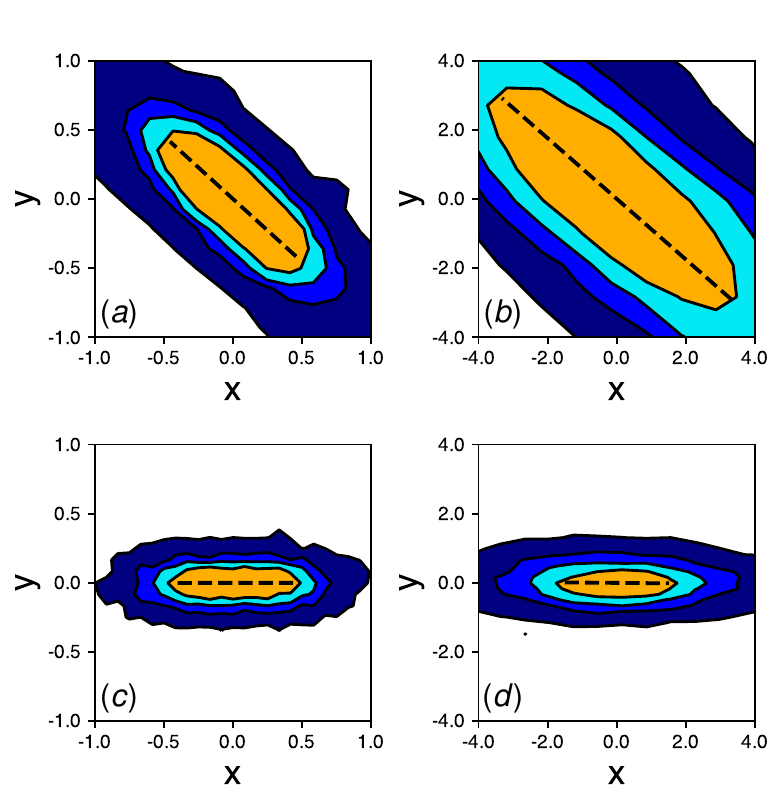}
  \caption{
    Contour plots of the average motion of an individual particle around its mean
    position in the steady state for cycles $n=1000$ to $n=2000$ plotted as
    a function of $y$ vs $x$ in units of the drive displacement $d$.
    The contours indicate areas in which each point has
    a minimum of 0.03\% (dark blue), 0.3\% (medium blue),
    0.75\% (light blue), and 1.5\% (orange) likelihood of being visited by the particle.
    The dashed lines,
    which are aligned with the drive in the overdamped limit and
    are at an angle $\theta \approx 41^\circ$ with respect to the drive in the skyrmion
    limit,
    indicate
    the average motion $\delta m_p$ along
    the first principal axis obtained via a principal
    component analysis.
    (a) Reversible motion in the skyrmion limit with
    $\alpha_m/\alpha_d=1$ and $d=14.3$,
    where $\delta m_p=1.2d$.
    (b) Irreversible motion in the skyrmion limit with
    $\alpha_m/\alpha_d=1$ and $d=29.9$,
    where $\delta m_p=8.8d$.
    (c) Reversible motion in the overdamped limit with
    $\alpha_m/\alpha_d=0$ and $d=14.3$,
    where $\delta m_p=0.87d$.
    (d) Irreversible motion in the overdamped limit with
    $\alpha_m/\alpha_d=0$ and $d=29.9$,
    where $\delta m_p=3.0d$.}
  \label{FIG_2}
\end{figure}

To understand the steady state behavior of the particles during each cycle,
we construct a contour plot of the average motion of an individual particle around
its mean position in the steady state.
For cycles $n=1000$ through $n=2000$, well outside the regime of transient behavior,
we sample the position of each particle
relative to its mean location every $0.1T$ time steps, and find the probability
that the particle will be observed at a given relative location, as plotted
in Fig.~\ref{FIG_2}.
The dashed line at the center of each panel of Fig.~\ref{FIG_2}
indicates the
average motion $\delta m_p$ along the first principal axis
obtained via principal component analysis.
For skyrmions with $\alpha_m/\alpha_d=1$,
shown in 
Fig. \ref{FIG_2}(a,b) for $d=14.30$ and $d=29.9$, respectively, the principal
axis is at an angle of $\theta\approx 41^\circ$ with respect to the drive 
direction.
This is very close to the
intrinsic skyrmion Hall angle
$\theta_{sk}^{\rm int}=
45^{\circ}$, as expected
based on
both simulations and experiments
that have previously demonstrated
a saturation of the skyrmion Hall angle
to the disorder free value
at large drives \cite{67,73,76,77,N,78}.
For the reversible state illustrated in Fig.~\ref{FIG_2}(a),
the particles move with the drive but return to
their starting positions after each
cycle, and therefore we expect the motion to be limited by the
drive displacement, $d$.
This is indeed the case, as indicated by the fact that
$\delta m_p \approx 1.2d$.
In contrast, in the irreversible state shown in Fig.~\ref{FIG_2}(b),
the particles
undergo an anisotropic random walk biased along the first
principal axis and do not return to their starting locations.
The average motion,
$\delta m_p \approx 8.8d$,
is much larger than in the reversible limit.
For overdamped particles with $\alpha_m/\alpha_d=0$,
the contour plots of the average motion in Fig.~\ref{FIG_2}(c,d)
for the reversible state at $d=14.3$
and the irreversible state at $d=29.9$
indicate that the overall motion is reduced compared to the skyrmion limit,
with $\delta m_p=0.87d$ in the reversible state and
$\delta m_p=3.0d$ in the irreversible state.

\begin{figure}
  \center
  \includegraphics[width=\columnwidth]{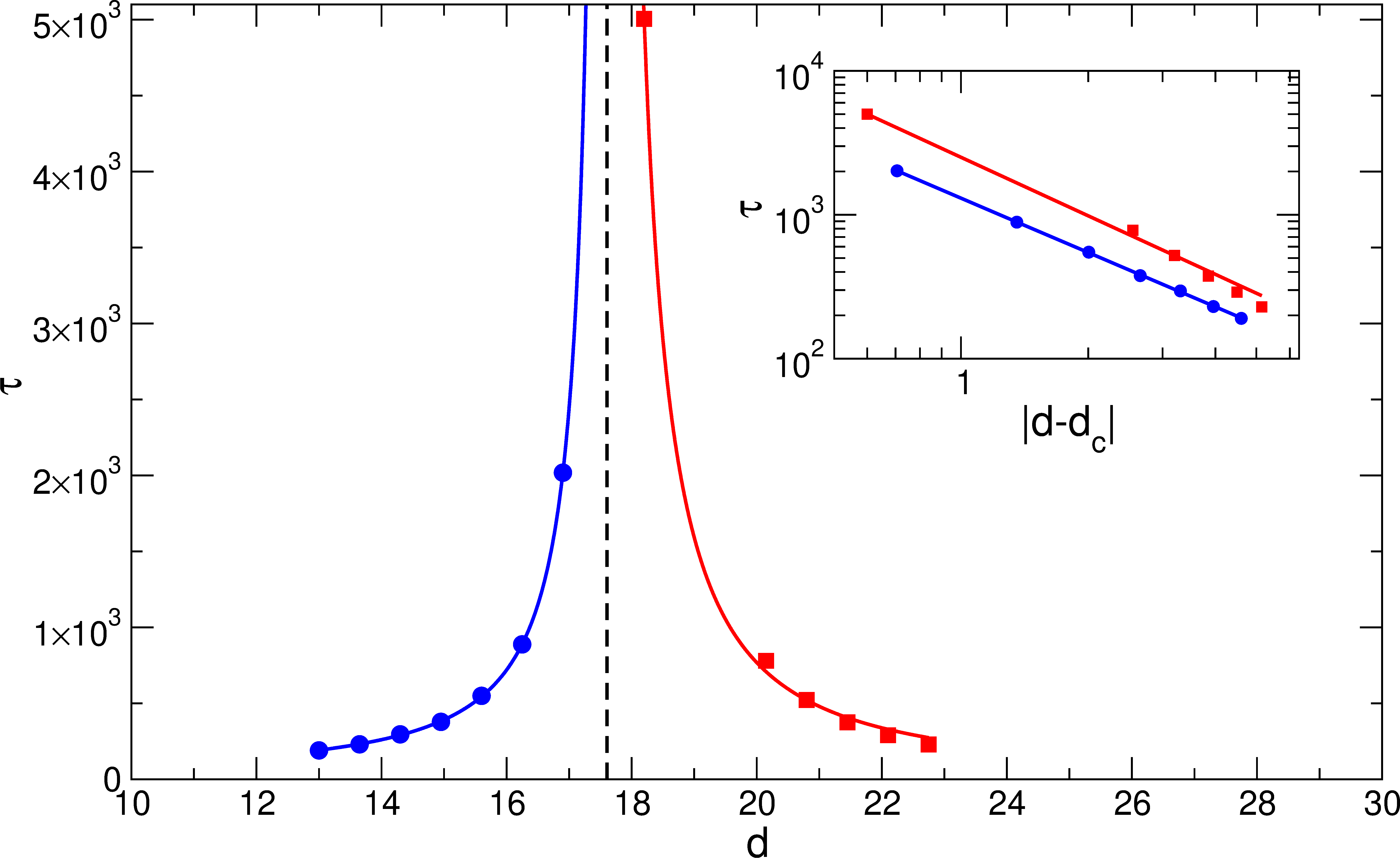}
  \caption{Characteristic time $\tau$ to reach steady state
    vs drive displacement $d$
    for the skyrmion system with $\alpha_m/\alpha_d=1$.
    A transition from
    reversible flow (blue circles) to irreversible flow (red squares)
    occurs at the critical displacement $d_{c}\approx 17.6$ (dashed line).
    The blue and red curves indicate power law fits to
    $\tau \propto |d-d_{c}|^{-\nu}$ with $\nu \approx 1.26$ in the reversible state and
    $\nu \approx 1.30$ in the irreversible state.
    Inset: The same data plotted as $\tau$ vs $|d-d_{c}|$ on a log-log scale.}
  \label{FIG_3}
\end{figure}

For the skyrmion system with $\alpha_m/\alpha_d=1$, we obtain the
characteristic time scale $\tau$ for a series of drive displacements $d$ by
fitting $F_{a}$ to Eq.~(\ref{EQ_4}).
In Fig.~\ref{FIG_3} we plot $\tau$ versus $d$ for both the reversible and
irreversible regimes, and find a divergence of $\tau$
at a critical
displacement $d_{c}\approx17.6$, 
consistent with a dynamical phase transition. 
On both sides of the transition, $\tau$
has the power law form $\tau \sim |d-d_{c}|^{-\nu}$, as shown in
the inset of Fig.~\ref{FIG_3},
with $\nu\approx 1.26$ in the reversible regime and
$\nu \approx 1.30$ in the irreversible regime.
These exponents are similar to the value
$\nu\approx 1.295$ expected for 2D
directed percolation \cite{12}
as well as to the value $\nu \approx 1.33$ observed
in the sheared colloid simulations of Ref.~\cite{11}.

\begin{figure}
  \center
  \includegraphics[width=\columnwidth]{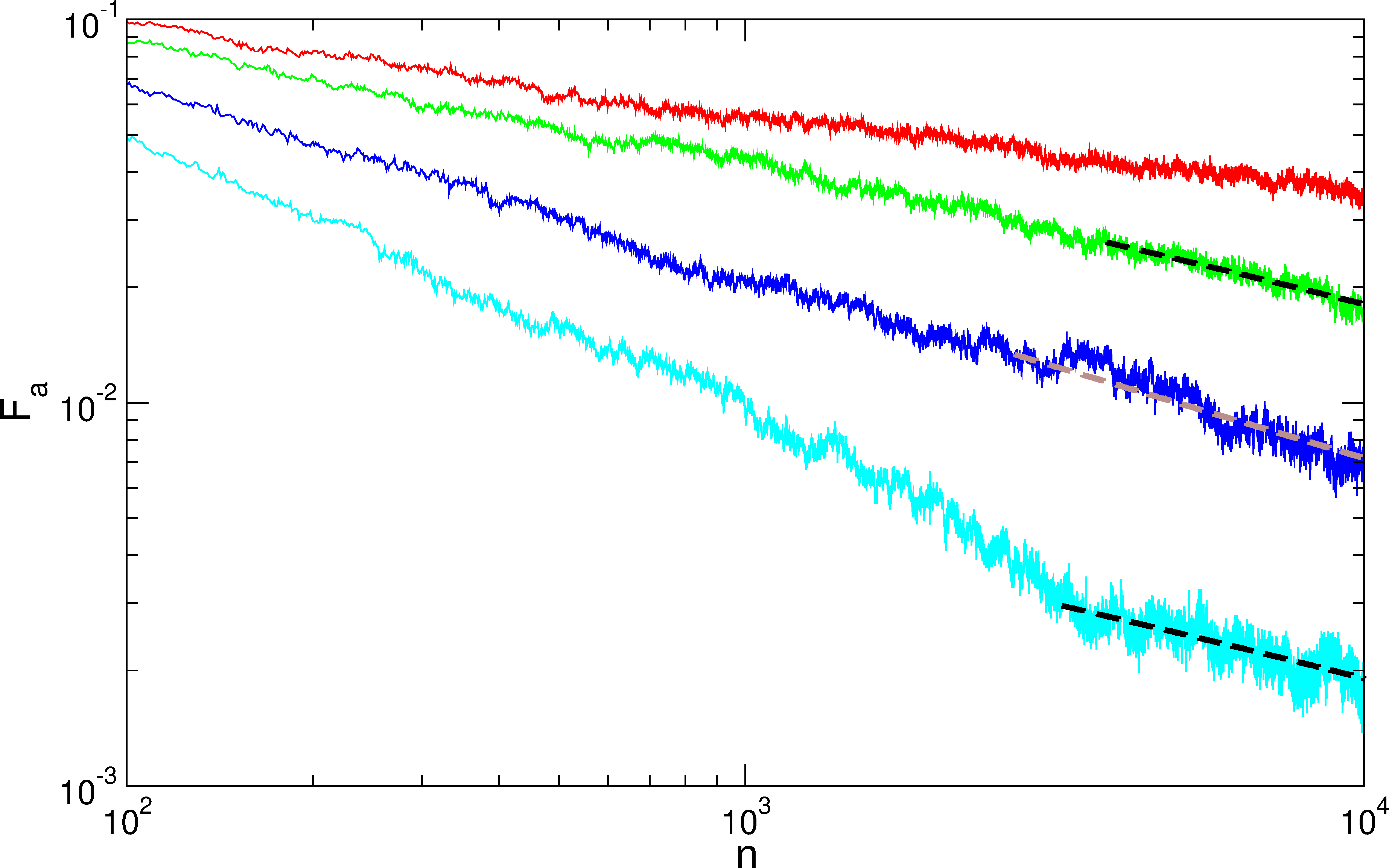}
  \caption{$F_a$, the fraction of active particles, vs $n$ for the skyrmion limit with
    $\alpha_m/\alpha_d=1$ for
    $d/d_c=1.07$, 1.03, 0.997, and 0.960, from top to bottom.
    The dashed lines are power law fits
    to
    $F_a \propto n^{-\alpha}$
    with $\alpha=0.32$ (black) and $\alpha=0.48$ (brown). }
  \label{FIG_4}
\end{figure}

In Fig. \ref{FIG_4} we plot 
the active fraction $F_a$ as a function of drive cycle $n$ in the
skyrmion system with $\alpha_m/\alpha_d=1$ near the transition
for drive displacements $d/d_c=0.960$ to 1.07.
At long times, we find a power law behavior with
$F_{a}\propto n^{-\alpha}$,
where $\alpha=0.48$ when
$d/d_{c}=0.997$.
This is similar to the expected exponents
$\alpha \approx 0.451$
for 2D directed percolation
and $\alpha=0.5$
for 2D conserved directed percolation
\cite{22}.

\begin{figure}
  \center
 \includegraphics[width=\columnwidth]{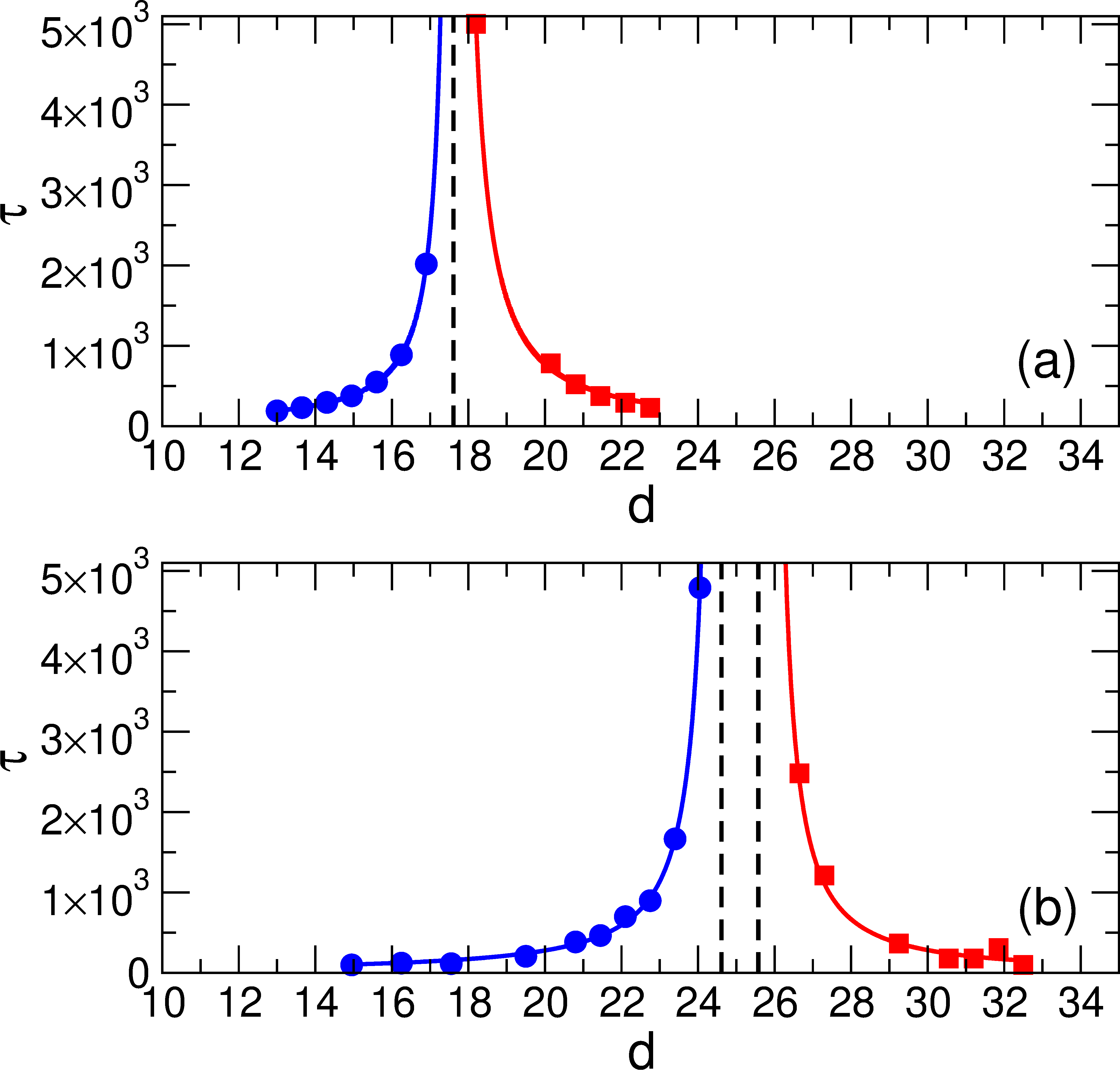}
 \caption{
   $\tau$ vs $d$ in the reversible (blue circles) and irreversible (red squares) regimes,
   along with the critical value $d=d_c$ (dashed lines) and
   power law fits to $\tau \propto |d-d_c|^{-\nu}$ (solid lines).
   (a) The skyrmion system with $\alpha_m/\alpha_d=1$, already shown in
   Fig.~\ref{FIG_3}, has $d_c=17.6$.
   (b) The overdamped
   system with $\alpha_m/\alpha_d=0$ has
   $d_c=24.6$ with $\nu=1.33$ in the reversible regime,
   and $d_c=25.8$ with $\nu=1.31$ in the irreversible regime.
   }
\label{FIG_5}
\end{figure}

In Fig.~\ref{FIG_5}(a),
we replot the values of $\tau$ versus $d$ for
the skyrmion system from Fig.~\ref{FIG_3}
in order to compare them with the behavior 
of $\tau$ in the overdamped system, shown in Fig.~\ref{FIG_5}(b).
We find power law exponents of
$\nu=1.33$ in the reversible state and $\nu=1.31$ in the irreversible state
for the overdamped system, similar to what we observe for the
skyrmion system,
indicating that omission of the Magnus term does not
appear to change
the universality class of the transition. 
There is, however, a large change in the value of the critical displacement $d_c$,
which falls at $d_c \approx 17.6$ in the skyrmion system but shifts to
the much higher value $d_c \approx 25$ in the overdamped system.
The Magnus term was previously shown to
enhance the effect of an external time-dependent noise
\cite{81},
and the suppression of the reversible regime
that we observe when we include the Magnus term
suggests that the Magnus term also enhances the chaotic nature
of the motion of skyrmions over the quenched disorder,
making it more difficult for the system to reach a reversible configuration.

\begin{figure}
  \center
  \includegraphics[width=\columnwidth]{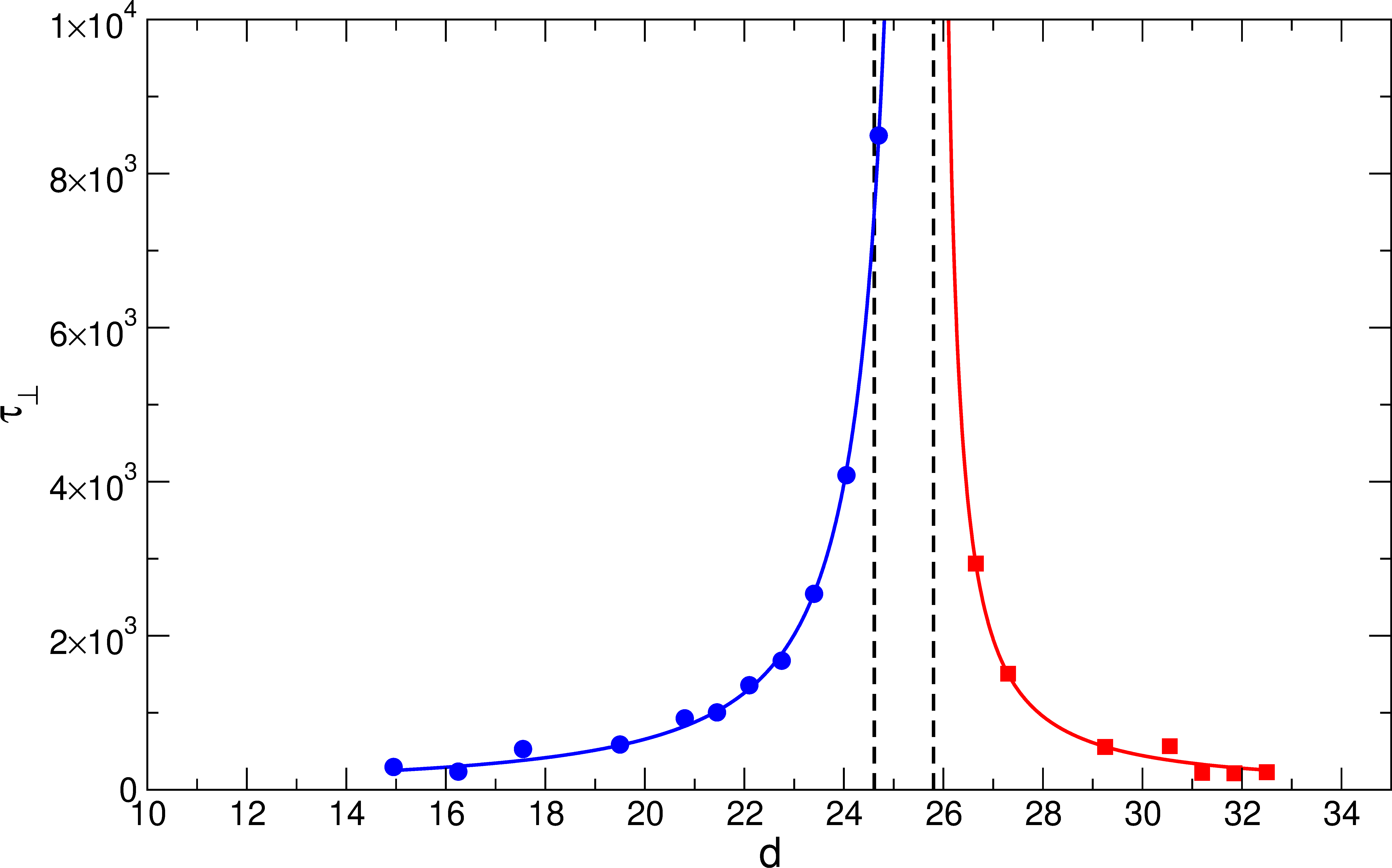}
  \caption{The characteristic time $\tau_{\perp}$ obtained from $R^2_{\perp}$, the
    cycle-to-cycle displacement transverse to the drive, vs $d$ for the
    overdamped system with $\alpha_m/\alpha_d=0$, along with
    solid lines showing fits to $\tau_{\perp} \propto |d-d_c^{\perp}|^{-\nu}$.
    The dashed lines indicate
    the two values of $d_c=24.6$ and $d_c=25.8$
    found in Fig.~\ref{FIG_5}(b) from fitting $\tau$ on either
    side of the transition.
    Here we find
    $d_{c}^{\perp}=25.8$ on both sides of the transition.}
\label{FIG_6}
\end{figure}

For the overdamped system in Fig.~\ref{FIG_5}(b),
we find that the value of $d_c$ is slightly different depending
on whether the transition is approached from the reversible or the irreversible
side.
In the reversible regime,
$d_{c}\approx 24.6$,
but in the irreversible regime,
$d_{c}\approx 25.8$.
This gap could 
indicate that there are two transitions instead of only one.
The first transition is
from fully reversible flow to a smectic state
in which the flow is reversible transverse to the drive
but irreversible in the direction of the drive.
The second transition is from this smectic state
to fully irreversible flow.
To test this idea,
we measure the cycle-to-cycle displacement in the transverse direction,
$R^{2}_{\perp}(n)=\langle(1/N)\sum_{i}^{N}[(\tilde{\bf R}_{i}(nT)-\tilde{\bf R}_{i}((n-1)T))\cdot\hat{\bf y}]^{2}\rangle$,
and obtain a characteristic time $\tau_{\perp}$ by fitting
$R^{2}_{\perp}$
to Eq.~(\ref{EQ_4}).
We plot $\tau_{\perp}$ versus $d$ on both sides of the transition in
Fig.~\ref{FIG_6}, where the dashed lines indicate the two values of $d_c$ obtained
by fitting the data in Fig.~\ref{FIG_5}.
By fitting $\tau_{\perp} \propto |d-d_{c}^{\perp}|^{-\nu}$, we find that
$d_{c}^{\perp}=25.8$ in both the reversible and irreversible regimes, indicating
that there is a single transition for motion perpendicular to the drive
at $d_c^{\perp}=25.8$.  The transition for motion parallel to the drive must then occur
at the lower value of $d_c^{||}=24.6$.  For $24.6 < d < 25.8$,
the flow is expected to follow
channels aligned with the driving direction.  This flow should be smectic in
nature, so that the
particles can slide past one another irreversibly
in the direction of the drive while remaining reversibly locked to a single
channel with no motion in the direction transverse to the drive.
It would be necessary to simulate much larger systems to clearly resolve the
behavior of the individual channels, which would be an interesting topic for a
future study.

\section{Summary}

We show that both skyrmions
and overdamped particles that are periodically driven over quenched disorder
undergo a transition as a function of drive period from a reversible state at small
drive periods, in which the particles return to the same position after each drive
cycle, to an irreversible state at large drive periods, in which the particle positions
gradually diffuse from cycle to cycle.
Near the transition, the fraction of active particles has a power law
time dependence
with an exponent $\alpha \approx 1/2$ in the skyrmion limit,
consistent with an absorbing phase transition.
The characteristic time required to reach a steady state
diverges as a power law at the transition with an exponent similar to that
expected for directed percolation for both the skyrmions and overdamped particles,
suggesting that inclusion of a Magnus term in the particle dynamics does not change
the universality class of the transition.
The Magnus term enhances the random motion generated by the quenched
disorder, and as a result the range of reversible behavior is much smaller for the
skyrmion system than for the overdamped particles.
We find evidence
that the overdamped system first transitions from the reversible regime to
a smectic state, and then undergoes a second transition from smectic flow to
fully irreversible flow as the drive period is increased.

\begin{acknowledgments}
We gratefully acknowledge the support of the U.S. Department of
Energy through the LANL/LDRD program for this work.
This work was carried out under the auspices of the 
NNSA of the 
U.S. DoE
at 
LANL
under Contract No.
DE-AC52-06NA25396 and through the LANL/LDRD program.
\end{acknowledgments}

\end{document}